\documentclass[a4paper,clock]{article}
\usepackage[dvips]{epsfig}
\usepackage{amssymb}
\usepackage{amsmath}
\usepackage{bm}
\usepackage{dcolumn}

\catcode`\@=11
\def\marginnote#1{}
\newcount\hour
\newcount\minute
\newtoks\amorpm
\hour=\time\divide\hour by60
\minute=\time{\multiply\hour by60 \global\advance\minute
by-\hour}\edef\standardtime{{\ifnum\hour<12
\global\amorpm={am}%
        \else\global\amorpm={pm}\advance\hour by-12 \fi
        \ifnum\hour=0 \hour=12 \fi
        \number\hour:\ifnum\minute<10
0\fi\number\minute\the\amorpm}}
\edef\militarytime{\number\hour:\ifnum\minute<10
0\fi\number\minute}

\def\draftlabel#1{{\@bsphack\if@filesw {\let\thepage\relax
   \xdef\@gtempa{\write\@auxout{\string
      \newlabel{#1}{{\@currentlabel}{\thepage}}}}}\@gtempa
   \if@nobreak \ifvmode\nobreak\fi\fi\fi\@esphack}
        \gdef\@eqnlabel{#1}}
\def\@eqnlabel{}
\def\@vacuum{}
\def\draftmarginnote#1{\marginpar{\raggedright\scriptsize\tt#1}}
\def\draft{\oddsidemargin -.5truein
        \def\@oddfoot{\sl preliminary draft \hfil
        \rm\thepage\hfil\sl\today\quad\militarytime}
        \let\@evenfoot\@oddfoot \overfullrule 3pt
        \let\label=\draftlabel
        \let\marginnote=\draftmarginnote

\def\@eqnnum{(\theequation)\rlap{\kern\marginparsep\tt\@eqnlabel}%
\global\let\@eqnlabel\@vacuum}  }

\def\numberbysection{\@addtoreset{equation}{section}
        \def\theequation{\thesection.\arabic{equation}}}

\def\underline#1{\relax\ifmmode\@@underline#1\else
 $\@@underline{\hbox{#1}}$\relax\fi}

\catcode`@=12
\relax

\numberbysection

\advance \topmargin by -\headheight
\advance \topmargin by -\headsep

\evensidemargin \oddsidemargin
\def\br{\begin{eqnarray}}
\def\er{\end{eqnarray}}
\def\be{\begin{equation}}
\def\ee{\end{equation}}

\def\({\left(}
\def\){\right)}

\relax

\def\a{\alpha}

\def\b{\beta}

\def\d{\delta}

\def\O{\Omega}

\def\pa{\partial}

\def\tp0{\Theta_{+}^{(0)}}
\def\tm0{\Theta_{-}^{(0)}}

\def\vp{\varphi}

\def\f#1#2#3 {f^{#1#2}_{#3}}

\def\win1{{\sf w_{1+\infty}}}

\def\Win1{{\sf W_{1+\infty}}}

\def\rlx{\relax\leavevmode}
\def\inbar{\vrule height1.5ex width.4pt depth0pt}
\def\IZ{\rlx\hbox{\sf Z\kern-.4em Z}}
\def\IR{\rlx\hbox{\rm I\kern-.18em R}}
\def\IT{\rlx\hbox{\rm I\kern-.18em T}}
\def\IC{\rlx\hbox{\,$\inbar\kern-.3em{\rm C}$}}
\def\IN{\rlx\hbox{\rm I\kern-.18em N}}
\def\IO{\rlx\hbox{\,$\inbar\kern-.3em{\rm O}$}}
\def\IP{\rlx\hbox{\rm I\kern-.18em P}}
\def\IQ{\rlx\hbox{\,$\inbar\kern-.3em{\rm Q}$}}
\def\IF{\rlx\hbox{\rm I\kern-.18em F}}
\def\IG{\rlx\hbox{\,$\inbar\kern-.3em{\rm G}$}}
\def\IH{\rlx\hbox{\rm I\kern-.18em H}}
\def\II{\rlx\hbox{\rm I\kern-.18em I}}
\def\IK{\rlx\hbox{\rm I\kern-.18em K}}
\def\IL{\rlx\hbox{\rm I\kern-.18em L}}
\def\one{\hbox{{1}\kern-.25em\hbox{l}}}
\def\0#1{\relax\ifmmode\mathaccent"7017{#1}%
B        \else\accent23#1\relax\fi}

\def\PRL#1#2#3{{\sl Phys. Rev. Lett.} {\bf#1} (#2) #3}

\def\PRB#1#2#3{{\sl Phys. Rev.} {\bf B#1} (#2) #3}

\def\RMP#1#2#3{{\sl Rev. Mod. Phys.} {\bf #1} (#2) #3}

\def\JPAMT#1#2#3{{\sl J. Physics A: Math. Theor.} {\bf A#1} (#2) #3}

\def\JPSJ#1#2#3{{\sl J. Phys. Soc. Japan} {\bf #1} (#2) #3}

\def\JHEP#1#2#3{{\sl JHEP} {\bf #1} (#2) #3}

\def\JNS#1#2#3{{\sl J. Nonlinear Sci} {\bf #1} (#2) #3}

\def\ADE#1#2#3{{\sl Advances in Differential Equations} {\bf #1} (#2) #3}
\def\AA#1#2#3{{\sl Applicable Analysis} {\bf #1} (#2) #3}
\def\SC#1#2#3{{\sl Science} {\bf #1} (#2) #3}
\def\DAO#1#2#3{{\sl Dynamics of Atmospheres and Oceans} {\bf #1} (#2) #3}
\def\PRR#1#2#3{{\sl Physical Review Research} {\bf #1} (#2) #3}
\def\JFM#1#2#3{{\sl Journal  of Fluid Mechanics} {\bf #1} (#2) #3}
\def\IJMCE#1#2#3{{\sl International Journal of Mathematics and Computer in Engineering} {\bf #1} (#2) #3}
\def\EJMBF#1#2#3{{\sl European Journal of Mechanics/B Fluids} {\bf #1} (#2) #3}
\def\RMP#1#2#3{{\sl Reviews in Mathematical Physics} {\bf #1} (#2) #3}
\def\CP#1#2#3{{\sl Communication Physics} {\bf #1} (#2) #3} 
 
\begin{document}

\begin{titlepage}

\vspace{.2in}
\begin{center}
{\large\bf Zero mode-soliton duality and pKdV kinks in Boussinesq system for non-linear shallow water waves}
\end{center}

\vspace{.2in}

\begin{center}

H. Blas$^{(a)}$, Ronal A. DeLaCruz-Araujo$^{(b)}$, 
N. I. Reynaldo Jr.$^{(a)}$, N. Santos$^{(a)}$, S. Tech$^{(a)}$ and  H.E.P. Cardoso$^{(a)}$
\par \vskip .2in \noindent

$^{(a)}$Instituto de F\'{\i}sica\\
Universidade Federal de Mato Grosso\\
Av. Fernando Correa, $N^{0}$ \, 2367\\
Bairro Boa Esperan\c ca, Cep 78060-900, Cuiab\'a - MT - Brazil \\
$^{(b)}$ School of Civil Engineering\\
 Universidad Nacional Aut\'onoma de Tayacaja Daniel Hern\'andez Morillo\\
Tayacaja, Huancavelica, Per\'u.
\normalsize
\end{center}

\vspace{.3in}

\begin{abstract}
\vspace{.3in}
A Boussinesq system for a non-linear shallow water is considered. The nonlinear and topological effects are examined through an associated matrix spectral problem. It is shown an equivalence relationship between the bound states and topological soliton charge densities which resembles a formula of the Atiyah-Patodi-Singer-type index theorem. The zero mode components describe a topologically protected Kelvin wave of KdV-type and a novel Boussinesq-type field. We show that either the $(1+1)-$ dimensional pKdV kink or the Kelvin mode can be mapped to the bulk velocity potential in $(2+1)-$ dimensions. 
\end{abstract} 

\end{titlepage}

\section{Introduction}

For a long time, it has been known that the linearized shallow water equations possess chiral edge oscillations, known as coastal Kelvin waves propagating around land masses. Similarly, eastward propagating chiral waves have also been found in the equatorial shallow water equations (see e.g. \cite{tong1} and references therein). Recently, a number of remarkable similarities have been found between the shallow
water equations in their linearized form and topological phases of matter \cite{delplace}. Quantum mechanically, chiral edge modes are the evidence for a topological
phase of matter, see \cite{qi} and references therein. The equatorial Kelvin waves of the classical shallow water equations were shown to belong to a class of classical topologically protected modes similar to those found in a topological insulator \cite{delplace, delplace1, delplace2}. 

In recent years, there has been an increasing interest in the study of topological insulators within the realm of condensed matter physics. These materials exhibit a fascinating phenomenon wherein the emergence of topological edge states serves as a hallmark of their unique topological phase. This characteristic is further underscored by the bulk-boundary correspondence, which establishes a fundamental relationship between the bulk properties of the material and the behavior of its edges states, see e.g. \cite{shapiro} and references therein.

Thus far, much of the investigation into topological physics has centered around the analysis of linearized versions of the systems under scrutiny. However, delving into the full nonlinear systems presents an intriguing and complex problem. Understanding the topological properties within these nonlinear frameworks represents a significant frontier in the field, promising to unveil new insights and potentially uncover novel phenomena previously inaccessible within the linearized paradigm. As researchers venture into this uncharted territory, they aim to unravel the intricate interplay between topology and nonlinearity, paving the way for a deeper comprehension of the rich tapestry of topological phases in condensed matter systems. The
nonlinear extension of the bulk-boundary correspondence uncovered the existence of nonlinear topological phases that are adiabatically disconnected from the linear regime \cite{sone, hadad, smirnova, kotwal, sone1, ezawa}.

The intricate interplay between topology and nonlinearity manifests itself in the formation of a unique class of solitons known as ``bulk solitons." Unlike their counterparts, the nonlinear edge solitons, which are confined to the edges of materials, bulk solitons inhabit the interior or bulk of the material, nestled within the topological bandgap. These bulk solitons emerge at the interfaces induced by nonlinearity within the material's topology. They represent localized, stable structures that persist within the bulk topological bandgap, serving as compelling manifestations of the complex interplay between nonlinear dynamics and topological properties.

Studies examining the properties and behavior of bulk solitons shed light on the intricate mechanisms underlying their formation and stability. The work of researchers, such as those referenced in \cite{li}, contributes to our understanding of these phenomena, providing valuable insights into the fundamental nature of solitons within topological materials. So,  further investigation and analysis are desirable in order to uncover the full extent of the role played by bulk solitons in shaping the behavior and functionality of topological systems, with potential implications for various fields ranging from photonics to quantum computing.
   
Here we provide a matrix spectral description of a nonlinear shallow
water dynamics in the context of a Boussinesq system and the related potential KdV model (pKdV), such that the topological and nonlinear properties become more manifest.  At the formal and calculational levels, the zero mode chiral waves coupled to the pKdV-type topological solitons are strikingly reminiscent of the chiral fermion zero modes of the Dirac equation coupled to the Toda field \cite{jhep22}. 
 
In the soliton sector, we show the  dependence of the $2+1$ dimensional bulk velocity potential on the pKdV kink. At the fluid-ar interface it is shown an equivalence relationship between the bound states and topological charge densities which resembles a formula of the Atiyah-Patodi-Singer-type index theorem. Moreover, it is shown that the Kelvin-like modes become the zero modes of the spectral problem. We will construct a duality mapping between the  topological solitons of pKdV-type and the zero mode bound states, and then, a mapping from either the kink or the zero mode state to the bulk velocity potential can be constructed.
 
Next section presents a Boussinesq system and pKdV equation. The bulk velocity potential is related to the pKdV kink. In sec. \ref{sec:matrix} a related matrix spectral problem is discussed. Soliton/zero mode duality and the future prospects are commented in sec. \ref{sec:discussion}. The Appendix \ref{app:pkdv} presents the kink solutions. 
 
\section{A Boussinesq system and KdV equation}
\label{sec:sim}
Let us consider a perfect fluid, i.e. incompressible and inviscid. So, one has the Euler equations of motion 
\br
\label{eu1}
\frac{\partial \bold{u}}{\partial t} + (\bold{u}.\nabla) \bold{u} = -\frac{\nabla p}{\rho} - g \hat{\bold{y}},
\er
such that $\nabla. \bold{u}  =0$. With $\bold{u} $, $\rho$, $p$, and $g$ being the velocity field, fluid density, pressure and acceleration constant, respectively. The unit vector $\hat{\bold{y}}$ points vertically upwards in the $y-$axis direction. We consider a two-dimensional fluid in cartesian coordinates $(x , y)$, where the coordinate $x$ denotes the horizontal axis. The Fig. 1 shows the Cartesian coordinates $(x,y)$ for the bi-dimensional fluid. The flat bottom is at $y=0$, $h_0$ is the depth of the fluid at rest, the wave amplitude is $a$, the solitary wave profile is represented by  $\eta(x,t)$, $l$ is the characteristic width of the pulse,  and $y_m$ the distance to the bottom at which some quantities can be measured. The small parameters  are defined as 
\br
\a \equiv \frac{a}{h_0},\,\,\,\, \b \equiv (\frac{h_0}{l})^2.
\er
\begin{figure}
\centering
\label{fig1}
\includegraphics[width=2cm,scale=2, angle=0,height=7cm]{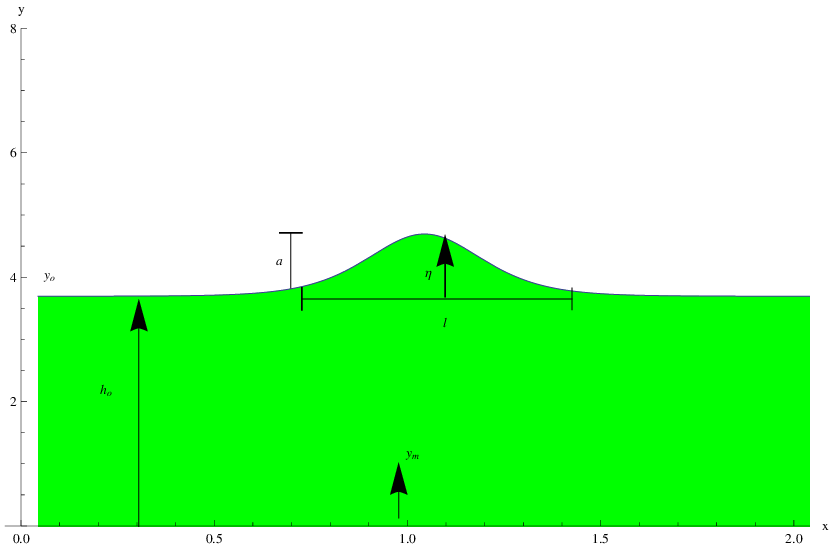} 
\begin{center}
\parbox{6in}{\caption{The plot shows the Cartesian coordinates $(x,y)$ for the bi-dimensional fluid . The bottom is at $y=0$, $h_0$ is the depth of the fluid at rest, the wave amplitude is $a$, the solitary wave profile is represented by  $\eta(x,t)$, $l$ is the characteristic width of the pulse,  and $y_m$ the distance to the bottom at which some quantities can be measured. }}
\end{center}
\end{figure}  

We will consider the dynamics of a traveling wave with amplitude $\eta$ on the free surface of an irrotational fluid. So, one can define the velocity potential as $\bold{u} = \nabla \psi$ and rewrite the equation (\ref{eu1}) as
\br
\left.\begin{array}{l}
\b \psi_{xx} + \psi_{yy} =0\\
	\frac{1}{\a}  (p-1) = -\psi_t - \frac{\a}{2}  \psi_x^2  - \frac{\a}{2\b} \psi_y^2  -  \frac{1}{\a} (y-1)  \end{array} \right\} \,\,\,\,  \,\,\,\,\,\,  0 < y < 1 + \a \, \eta \label{bulk1},
\er
in the interior of the fluid and in adimensional variables. One considers the following boundary conditions 
\br
\label{kindyn1}
\left.\begin{array}{l}
\eta_{t} + \a \psi_x  \eta_x  = \frac{1}{\b} \psi_y, \\
\psi_t +  \frac{\a}{2}  \psi_x^2  + \frac{\a}{2\b} \psi_y^2   +  \eta = 0\end{array} \right\}\,\,\,\,\,  \,\,\,\,\,\,y = 1 + \a \, \eta,
\er
and
\br
\label{bc2i}
\psi_y =0   \,\,\,\,\,  \mbox{for} \,\,\,\,\,\,    y = 0.
\er
 
The system (\ref{bulk1}) is subject to the boundary conditions (\ref{kindyn1})-(\ref{bc2i}), which are defined for the unknown wave profile $\eta$. Ordering of two small parameters in the shallow water
wave problem and a systematic procedure for deriving an equation for
surface elevation for a prescribed relation between the orders of the two
expansion parameters $\a$  and  $\b$ has been developed in \cite{burde}. Here we follow the standard approach with small parameters $\a$ and $\b$ of the same order. For an earlier pedagogical account, see e.g. \cite{larosa1} and references therein. One has
\br
\label{exp1}
\psi(x,y,t)  = \vp_{0}(x,t) -\frac{\b}{2} y^2 \vp_{0 xx}(x,t)+ \frac{\b^2}{24} y^4 \vp_{0 xxxx}(x,t) - \frac{\b^3}{720} y^6 \vp_{0 xxxxxx}(x,t) + {\cal O}(\b^4),
\er
where $ {\cal O}(\b^4)$ stands for the terms of order higher or equal to $\b^4$. Defining the Boussinesq field as $w \equiv  \vp_{0x}$, from the kinematic boundary condition in (\ref{kindyn1}) one can write 
\br
\label{bouss1}
\eta_t + [(1+ \a \eta) w]_x - \frac{1}{6} \b w_{xxx} + {\cal O}(\a^2,\b^2,\a \b) =0.
\er
Likewise,  from the dynamic boundary condition in (\ref{kindyn1}), upon taking its $x-$derivative, one has 
\br
\label{bouss2}
\eta_x + w_t + \a w w_x - \frac{1}{2} \b w_{xxt} +  {\cal O}(\a^2,\b^2,\a \b) =0.
\er
We will regard the set of eqs.  (\ref{bouss1})-(\ref{bouss2}) as a Boussinesq system of equations. For a wide class of Boussinesq-type systems, see e.g. \cite{alazman, chen, bona1}.  Up to the first order in $\a$ and $\b$, one has 
\br
\label{ww2}
w = \eta - \frac{1}{4} \a \eta^2   +  \b \frac{1}{3}  \eta_{xx}, 
\er
and the following non-linear differential equation for the field $\eta$
\br
\label{kdv0}
\eta_t + \eta_x + \frac{3}{2} \a \eta \eta_x + \frac{1}{6} \b  \eta_{xxx} =0. 
\er
This is just the well known KdV equation. Let us analyze  the dispersion relations of the linearized versions of the Boussinesq system (\ref{bouss1})-(\ref{bouss2}) and the KdV model, respectively. So, for (\ref{bouss1})-(\ref{bouss2}) one has
\br
\label{Bomega}
\omega^{\pm}_1 = \pm \frac{k \sqrt{1+ \frac{1}{6}\b k^2}}{\sqrt{1+ \frac{1}{2}\b k^2}}.
\er
The phase and group velocities are $\omega^{\pm}_1/k  = \pm \frac{\sqrt{1+ \frac{1}{6}\b k^2}}{\sqrt{1+ \frac{1}{2}\b k^2}}$ and $\frac{d \omega_1^{\pm}}{dk}  = \pm \frac{1}{\sqrt{3}} \frac{8+(2+\b k^2)^2}{(2+\b k^2)^{3/2}\sqrt{6+\b k^2}}$, respectively, which are bounded for large enough $k$. For the linearized KdV equation the dispersion relation becomes
\br
\label{om0}
\omega = \frac{1}{6} \( 6 k - \b k^3\), 
\er 
which implies the phase velocity  $\omega/k  = 1- \frac{\b}{6} k^2$ and group velocity $\frac{d \omega}{dk}  = 1- \frac{\b}{2} k^2$, which are unbounded from below for large enough $k$. In the Fig. 2  we plot $\omega_1^{\pm}$ (red/blue) and $\omega$ (dashed), respectively, for $\b =0.01$. Notice the linear behavior and overlap of $\omega_1^{+}$ and $\omega$ ($\omega_1^{+} \approx \omega \approx k$)  for large wavelength. Note that the long wavelength
(or shallowness condition) parameter $\b$ is considered to be a small parameter, i.e. $\b << 1$; so, for small $k$ (large wavelength) the linear behavior of the both $\omega$  and $\omega_1^{+} $ is maintained for small $\b$. Moreover, one can notice that qualitatively the plots in Fig. 2 do not alter significantly for different values of the small parameter $\b$.   
\begin{figure}
\centering
\label{fig2}
\includegraphics[width=2cm,scale=2, angle=0,height=6cm]{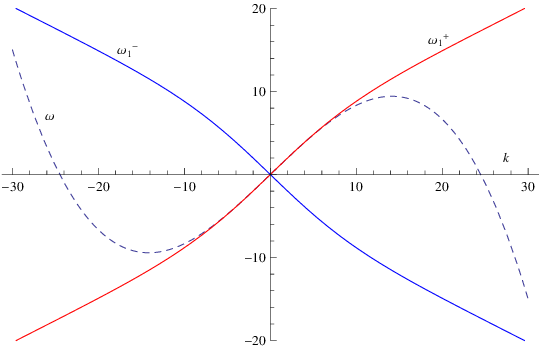} 
\begin{center}
\parbox{6in}{\caption{Dispersion relations for Boussinesq $\omega_1^{\pm}$ (red/blue) and KdV $\omega$ (dashed) models. Boussinesq and KdV are gapless. Note their linear behavior around the origin.}}
\end{center}
\end{figure}

\subsection{Bulk velocity potential and interface pKdV kink}   

Here we establish a relationship between the bulk velocity potential  and the  interface fields. The properties of the fluid-air interface are encoded into the Boussinesq $w$ and KdV $\eta$ fields of the system (\ref{bouss1})-(\ref{bouss2}). Up to the first order in $\a$ and $\b$ it will be sufficient to consider the equivalent description provided by the relationship  (\ref{ww2}) and the KdV equation itself (\ref{kdv0}). Let us consider the nonlinear soliton model dubbed as the potential KdV model (pKdV) defined for the field $q$ ($\eta =\pa_x q $). So, from (\ref{kdv0}) one can get 
\br
\label{pkdv}
q_t + q_x + \frac{3}{4} \a  q_x^2 + \frac{1}{6} \b q_{xxx} =0,
\er
where an overall $x-$derivative has been removed. In order to find the topological soliton of interest we will consider the eq.  (\ref{pkdv}) with nonvanishing boundary conditions (see Appendix).  
 
Replacing $w = \vp_{0x}$ and  $\eta =\pa_x q$ into  (\ref{ww2}) and  integrating in $x-$coordinate once, one can get
\br
\label{vpq}
\vp_0 = q + \frac{\b}{3} q_{xx} - \frac{\a}{4} \int_{-\infty}^{x} dx \, (q_{x})^2. 
\er
Notice the appearance of a non-local contribution in the last term. On the other hand,  the velocity potential $\psi(x,y,t)$ in (\ref{exp1}) depends on powers of the variable $y$ and the $x-$derivatives of $\vp_0$. So, according to (\ref{vpq}), the  $\b^{0}$ order of $\psi(x,y,t)$ in (\ref{exp1}) gets a non-local contribution of order $\a$.

Remarkably, the velocity potential $x-$dependence is carried out only by  the $\vp_0(x)$ field and its higher order $x-$derivatives in (\ref{exp1}); so, the potential $\psi(x,y,t)$ at every point $(x,y)$ can be reconstructed by knowing the field $\vp_0(x)$ or, alternatively, the pKdV field $q$ according to (\ref{vpq}), since the even powers of the coordinate $y$ appears explicitly in the expansion of $\psi(x,y,t)$ in (\ref{exp1}).   

In order to see the topological and non-local contribution to the potential $\psi(x,y,t)$ let us write its expression explicitly  in terms of the pKdV field $q$ as
\br
\nonumber
\psi(x,y,t) & =& q(x) - \frac{\a}{4} \int_{-\infty}^{x} dx \, (q_{x})^{2} + \frac{\b}{6} (2-3y^2) q_{2x}(x) -\frac{\b^2}{24} y^2 (4-y^2) q_{4x}(x) +  \\ &&\frac{\b^3}{720} y^4 (10-y^2) q_{6x}(x) -  \frac{\b^4}{2160} y^6 q_{8x}(x) + \frac{\a \b}{4} y^2 q_{x} q_{2x} - \frac{\b^2 \a}{48} y^4 ( 3q_{2x} q_{3x} + q_x q_{4x})+\nonumber 
\\
&&\frac{\b^3 \a}{1440} y^6 (10 q_{3x} q_{4x} + 5 q_{2x} q_{5x} + q_x q_{6x}),
 \label{exp1q}
\er
where $q_{nx}$ denotes the $n^{th}$ order $x-$derivative of the field $q$.

Consider a kink solution $q_k$ of the pKdV model (\ref{pkdv}) by setting $a=b=0$ in (\ref{qxt}). Denote the velocity components as $\bold{u}\equiv\nabla \psi = \(U_1 \,, \,U_2 \)$. Remarkably, one can write $U_{1,2}$ as functionals  of the kink $q_k$ as 
\br\nonumber 
U_1 &=& \frac{1}{12} A_0 k[ -3 A_0 k \a + 4 (3 - k^2(2- 3 y^2)\b)] - \frac{k}{6 A_0}[6 - 3 A_0 k \a - 8 k^2 (2-3y^2)\b] (q_k-q_0)^2-\\
&& \frac{1}{4 A_0^3} k^2 [A_0 \a + 4 k (2-3 y^2) \b] (q_k-q_0)^4.\label{v1q}\\
U_2 &= & \frac{2 \b k^2 y}{A_0^2} (q_k - q_0) [A_0^2 - (q_k - q_0)^2].\label{v2q}.
\er
So, one can argue that the velocity components in the bulk carry the topological information of the pKdV kink  $q_k$. 
\begin{figure}
\centering
\label{fig3}
\includegraphics[width=2cm,scale=2, angle=0,height=4cm]{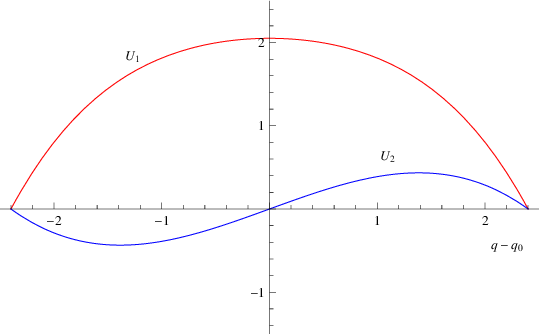} 
\begin{center}
\parbox{6in}{\caption{The velocity components $U_1$ (red) and $U_2$ (blue) of (\ref{v1q})-(\ref{v2q}) at $y= \sqrt{\frac{2}{3}}$ as  functionals of the  pKdV kink $(q-q_0) \in [-A_0, A_0]$ for  $q_0=0, a=0, k =1.2, \a = 0.4, \b = 0.2, A_0 = 2.4$.}}
\end{center}
\end{figure}
The Fig. 3 shows the velocity components $\(U_1(q-q_0)\, ,\, U_2(q-q_0)\)$ in (\ref{v1q})-(\ref{v2q}) as functions of the topological  pKdV kink $(q_k-q_0)$ for $y = y_m = \sqrt{\frac{2}{3}}$. It shows the symmetric and anti-symmetric properties of $U_1$ and $U_2$, respectively, as functionals of the kink $(q-q_0)$, as it interpolates from $-A_0$ to $+A_0$.  Note that the kink asymptotically becomes $[q(x = \pm \infty) - q_0] \Rightarrow \pm  A_0$, therefore, at the center of the kink $q= q_0$ (around the origin of the plots in Fig. 3) the velocity $x-$component $U_1(q-q_0=0)$ assumes its maximum value, whereas the vertical $y-$component $U_2(q-q_0 =0)$ vanishes.  As the kink becomes asymptotically constant $[q(x = \pm \infty) - q_0] \rightarrow  \pm A_0$ the both components $U_{1,2}$ tend to zero.

\section{Matrix spectral problem and kink/bound state correspondence}   
 \label{sec:matrix}

In this section we uncover the relevant composite fields which describe the interplay between the non-linearity and topology of the model.  So, we consider the nonlinear effects of the Euler equations represented by the Boussinesq system (\ref{bouss1})-(\ref{bouss2}).  Recently, some authors have linearized the shallow water wave equations in order to study the role played by topology in the generation of the equatorial Kelvin and Yanai waves \cite{delplace1, delplace2}. Here, instead, we consider the full system of nonlinear equations of motion  (\ref{ww2})-(\ref{kdv0}) in order to analyze the topological properties of the model. 

In order to examine the non-linear and topological degrees of freedom let us rewrite the  eqs.  (\ref{ww2})-(\ref{kdv0}) as 
\br
\label{chi110}
  \pa_x \chi_{10} + B_2 \chi_{10} - A_2 \chi_{20}   &=& 0\\
\label{chi220}
  \pa_x \chi_{20}  - B_1 \chi_{10} - B_2 \chi_{20}   &=& 0,
\er
with the definitions
\br
\chi_{10} &\equiv& \sqrt{w}, \,\,\,\chi_{20} \equiv \frac{\sqrt{\a}}{2} \eta, \label{chi120}\\
 B_1 &\equiv& - \sqrt{\a} [w^{1/2}]_x  ,\,\,\,\,B_2 \equiv - \frac{7}{2} \a \eta_x \label{m111o},\,\,\,\, A_2 \equiv  \frac{1}{\sqrt{\a}} \frac{w_x - 7 \a w \eta_x}{\sqrt{w} \eta},\label{B12i}
\er
where $w$ and $\eta$ are the Boussinesq and KdV fields, respectively. In fact, the eq. (\ref{chi220}), upon substitution of the relationships (\ref{chi120}), and taking into account (\ref{B12i}), become 
\br
\label{kdvso}
\eta_x + \frac{3}{2} \a \eta \eta_x + \frac{\b}{6} \eta_{xxx} =0.
\er
This is just the static version of the KdV eq. (\ref{kdv0}). Moreover, the relationship (\ref{chi110}) is an identity provided that  the static version of (\ref{ww2})-(\ref{kdv0}) are taken into account. 

Moreover, from  (\ref{ww2}) and (\ref{chi120}) one can write the suggestive equation
\br
\label{chi12o}
\chi_{10}^2 + \chi_{20}^2 &=&  \eta +  \b \frac{1}{3}  \eta_{xx}  \\
&\equiv &    \pa_x Q_o \label{chi12qo}.\er
 
Since the potential KdV field $q$ is defined as $\eta = \pa_x q$, the field $Q_o$ can be written as
\br
\label{Qoqk}
Q_o &\equiv & q  + \frac{\b}{3} q_{xx} + const.  \label{Q0}
\er
The KdV soliton $\eta$, the Boussinesq field $w$, kinks $\{q, Q_o\}$, and bound states $\{\chi_{10},\chi_{20}\}$ are shown in Fig. 4 (see Appendix \ref{app:pkdv} and set  $a=q_0=0$). Notice that the topological soliton profiles (kinks $\{q, Q_o\}$) differ slightly. The bound state solitons $\{\chi_{10},\chi_{20}\}$ represent the bulk solitons in the one-dimensional air-fluid interface, whereas the field $Q_o$ represents a topological kink, such that the bound states are localized inside the kink, as shown in the Fig. 4.
     
\begin{figure}
\centering
\label{fig4}
\includegraphics[width=2cm,scale=2, angle=0,height=5cm]{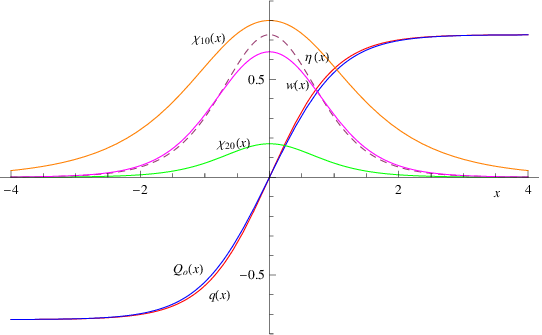} 
\begin{center}
\parbox{6in}{\caption{KdV soliton $\eta$ (dashed), bound states $\{\chi_{10},\,\chi_{20}\}$ (orange and green), pKdV kink $q$ (red), kink $Q_o$ (blue) and Boussinesq field $w$ (magenta) for $a=0, \d_o=0, k =1, \a = 0.2, \b = 0.1$. Note that the bound state solitons are localized inside the central region of the kink  $Q_o$.}}
\end{center}
\end{figure} 

It is assumed that $Q_o$ possesses nontrivial topological properties allowing the existence of a topological charge. In fact, from (\ref{chi12o})-(\ref{chi12qo}) one has
\br 
\label{charge0}
\int_{-\infty}^{+\infty} \, dx \, (\chi_{10}^2 + \chi_{20}^2) &=& Q_o(+\infty) - Q_o(-\infty).
\er 
So, the quantity $\int_{-\infty}^{+\infty} \, dx \, (\chi_{10}^2 + \chi_{20}^2)$ is a topological invariant. One can regard the $Q_o$ field as a topological soliton and the fields $\chi_{1,2}$ as the components of a bound state. 
A considerable amount of work on topological phases has emerged relevant to a variety of physical systems. In this context, an important concept is the bulk-boundary correspondence, such that the topological invariant related to the
bulk states counts the number of gapless boundary states. So, topological phenomena can be related to boundary phenomena \cite{kunst}. Originally found in condensed matter physics, robust edge states are found in practically all wave dynamics. 

Moreover, one can show a duality correspondence in the interface between a topological soliton $Q_{kink}$ related to the pKdV model and certain bound state configurations $\chi_{1,2}$ related to the Boussinesq system. Therefore,  (\ref{chi12o})-(\ref{chi12qo})  is analog to a formula of the Atiyah-Patodi-Singer-type relating the bound state and topological charge densities; e.g. as studied in the Toda system coupled to the Dirac field \cite{jhep22}. In this context, this phenomenon can be associated to a gapless state formed by a zero mode bound state solution coupled to a topological kink-type soliton.

In this context one can generalize the system (\ref{chi110})-(\ref{chi220}) to the following spectral problem
\br
\label{chi11}
  \chi_{1x} + B_2 \chi_{1} - A_2 \chi_{2} + \O \chi_2 &=& 0\\
\label{chi22}
  \chi_{2x}  - B_1 \chi_{1} - B_2 \chi_{2} - \O  \chi_1 &=& 0,
\er
with $\Omega$ the spectral parameter, $\chi_{1,2}$ some Dirac-like fields and $B_1, B_2, A_2$ defined in (\ref{B12i}). One can assume the boundary conditions  
\br
\label{bc1}
\chi_{1,2} (|x|= + \infty)  &\rightarrow & 0, \\
\label{bc2}
A_2(|x|= + \infty)  & \rightarrow & A_{2o},\\
\label{bc3}
B_j(|x|= + \infty)  & \rightarrow & B_{jo},
\er
with $A_{2o}$  and $B_{jo}$ constants. In addition, one considers the first order differential  equation for the scalar $Q$
\br
\label{chi12w}
\chi_1^2 + \chi_2^2 &=& \pa_x Q,      
\er
such that $Q$ is a field representing a topological soliton, provided that the field components $\chi_{1,2}$ describe the relevant bound states. The system  (\ref{chi11})-(\ref{chi22}) and the relationship (\ref{chi12w}) are expected to arise for the higher order  Boussinesq system (\ref{bouss1})-(\ref{bouss2}), i.e. when the terms of order higher or equal to $ {\cal O}(\b^2, \a^2)$ are considered. Note that, in analogy to $Q_o$ in (\ref{chi12o})-(\ref{chi12qo}), the eq. (\ref{chi12w}) for the field $Q$ must be consistent with the relevant order extension of the eq. (\ref{ww2}) for the Boussinesq field $w$.  

So, the system of equations (\ref{chi11})-(\ref{chi22}) and (\ref{chi12w}) describes the dynamics of a Dirac-like field $\chi$ coupled to the scalar fields $w, \eta$. Then, one can examine  the bound states associated to the fields $\chi_{1,2}$ satisfying the above system of equations with b.c.  (\ref{bc1})-(\ref{bc3}). In this construction, the bound state fields $\chi_{10}, \chi_{20}$ in (\ref{chi120}) can be regarded as the zero-modes ($\O =0$) of the system of eqs.  (\ref{chi11})-(\ref{chi22}) for the first order   
Boussinesq system (\ref{bouss1})-(\ref{bouss2}). It is assumed that $Q$ in (\ref{chi12w}) possesses nontrivial topological properties allowing the existence of a topological charge. In fact, from (\ref{chi12w}) one has
\br 
\label{charge}
\int_{-\infty}^{+\infty} \, dx \, (\chi_1^2 + \chi_2^2) &=& Q(+\infty) - Q(-\infty).
\er 

The presence of non-linearity and the topological field $Q$ imply a gapped spectra of the system (\ref{chi11})-(\ref{chi22}) and consequently, the appearance of chiral zero modes. In analogy, it has been presented in \cite{jhep22} a kink-type Toda soliton coupled to Dirac spinor bound-states, and established a zero mode-soliton duality, as well as the existence of in-gap and boundary states in the continuum (BIC) bound states. In view of our results, the nonlinear shallow water equations furnish a new playground to study the  zero mode/soliton dualities. Topological ideas associated to chiral modes in fluids have recently been considered \cite{tong1, delplace}, and  it was shown a topological origin of the equatorial Kelvin and Yanai modes. 

Since the dynamics of  Kelvin waves in certain regimes can be deduced directly from the KdV model \cite{boyd1}, one can argue that the KdV-type bound state $\chi_{20}$ would describe a topologically protected Kelvin-type wave. To our knowledge, the $Q_o$ kink and the $\{\chi_{10},\chi_{20}\}$ bound states presented here are new. 

\subsection{The in-gap zero mode and threshold states}   

Let us examine the free wave states  associated to the system (\ref{chi11})-(\ref{chi22}). In order to linearize this system let us take some asymptotic values according to the boundary conditions (\ref{bc2})-(\ref{bc3}).  So, let us set  $B_{1o} = B_{2o}=0$ and $A_{2o}  \neq 0$. Then, by substituting $\chi_1 = \chi^{(o)}_{1} \cos{(k x)}$  and $\chi_2 = \chi_{2}^{(o)} \sin{(k x)}$ into the system (\ref{chi11})-(\ref{chi22}) one can get the dispersion relationships
\br
\Omega^{\pm} =\pm \frac{1}{2} (A_{2o} + \sqrt{k^2 + A_{2o}^2}),\,\,\,\, \Omega^{\pm}_o = \frac{1}{2} (\mp A_{2o} \pm \sqrt{k^2 + A_{2o}^2}),\,\,\,\,\,\, A_{2o}> 0.
\er
In Fig. 5  we plot these dispersion relations. Notice the appearance of the gap between the upper $\Omega^{+}$ and lower $\Omega^{-}$ red lines.
\begin{figure}
\centering
\label{fig5}
\includegraphics[width=2cm,scale=2, angle=0,height=6cm]{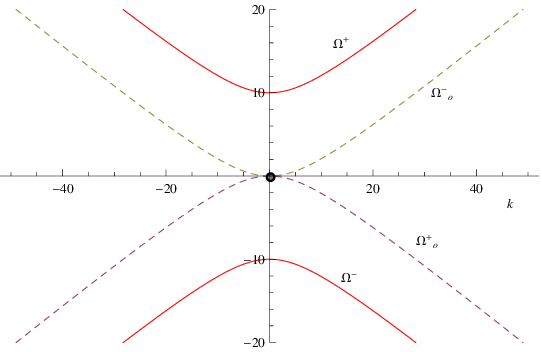} 
\begin{center}
\parbox{6in}{\caption{Dispersion relations for Dirac-type system $\{\Omega^{\pm},\Omega^{\pm}_o\}$ (red and dashed), for   $B_{1o} = B_{2o} = 0$ and $A_{2o} = 10$. Note that the Dirac-type model is gapped. The black dot at the origin shows the degenerated zero-mode states.}}
\end{center}
\end{figure}
One can argue that the Boussinesq dispersion relations $\omega_1^{\pm}$ (\ref{Bomega}) (Fig. 2 red/blue) split into the dispersion relations $\Omega^{\pm}$ (Fig. 5) which develop a gap due to the non-vanishing boundary condition (\ref{bc2}) of the  composite field $A_2$. The black dot in Fig. 5 corresponds to zero-modes and this degenerated state plays a central role in the topology of shallow water waves. In the nonlinear regime we have  seen above that the zero-modes $\chi_{10, 20}$ become dual to the topological soliton $Q_{o}$.

\section{Conclusions and discussions}
\label{sec:discussion}
We uncovered the fields which describe the interplay between the non-linearity and topology of the Boussinesq system (\ref{bouss1})-(\ref{bouss2}). It is considered the equivalent system of equations (\ref{ww2})-(\ref{kdv0}) in order to analyze the topological properties of the model. This is in contradistinction to the linearized shallow water wave equations which have recently been considered in order to study the role played by topology in the generation of the equatorial waves \cite{delplace1, delplace2}. The bound states $\chi_{1,2}$ and the topological soliton $Q$ satisfy the equivalence relationship (\ref{chi12w}); so, it can be regarded as the analog of the bulk-boundary correspondence in condensed matter systems \cite{kunst}. In the first order in $\a, \b$ system (\ref{bouss1})-(\ref{bouss2}) (or equivalently  (\ref{ww2})-(\ref{kdv0})), one has the zero modes $\{\chi_{01}, \chi_{02}\}$ and the equivalence soliton/bound state becomes (\ref{chi12o})-(\ref{chi12qo}) where the field $Q_o$ is related to the pKdV field $q$ through (\ref{Qoqk}). 

The pKdV field $q$ can be written in terms of the field $Q_o$ by solving the linear equation (\ref{Qoqk}), then $q$ will become a functional of $Q_o$. Moreover, since $w$ and $\eta$ can be written in terms of $Q_o$ the eqs. (\ref{chi120}) define a mapping between the kink $Q_o$ and the $\{\chi_{10},\chi_{20}\}$ zero modes in the soliton sector. So, one can regard the eqs. (\ref{chi120})  together with (\ref{chi12qo}) as a zero mode/soliton duality mapping. 

Notice that from (\ref{chi120}), using $\eta=\pa_x q$, one can get $ q = \frac{2}{\sqrt{\a}}\int^x dx'\, \chi_{20}$. So, one can write a mapping from either the kink $Q_o$ or the zero mode bound state $\chi_{20}$ to the bulk velocity potential $\psi(x,y,t)$ as defined in (\ref{exp1q}).  

It would also be interesting to study in more detail the non-linear soliton-soliton scattering processes and the duality correspondence discussed above in the context of the quasi-integrable KdV models \cite{jhep20}. The relevance of these ideas to the set of  Boussinesq-type systems and their related nonlinear shallow water wave equations deserve to be analyzed in the lines above. In particular, some extensions and generalizations of the KdV model have been appeared in the treatment of shallow water waves with non-planar bottom topography as boundary conditions, see e.g. \cite{cheng1} and references therein. 

Important topics scarcely covered in the recent literature on the so-called bulk-interface correspondence are the non-hermiticity and nonlinear effects on the topology, as discussed recently in \cite{nonher1} in the context of non-hermitian Toda models coupled to spinor fields. Another avenue worthy of investigation due its relevance in many  applications would be the fractional derivative modifications of the Boussinesq systems and their related pKdV-type equations in order to examine the bulk-interface relationship in that scenario. Recently, certain analytical methods have been proposed in order to deal with fractional derivative $\phi^4$ model with kink-type solitons and related pseudo-hyperbolic systems \cite{gasmi, sadeq}.  We will report on these important topics in the future.  
 
Finally, at this point one must distinguish between the true solitons of integrable systems, such as the bright and kink solitons of the integrable KdV and pKdV models, respectively, from the solitary waves like the kink of the non-integrable $\phi^4$ model. The true solitons of integrable systems undergo elastic collisions, and the only outcome of two-soliton collision being a phase shift. However, in the context of some deformations of integrable systems, the so-called quasi-integrable models, the behavior of their solitary waves resemble to the solitons of their undeformed counterparts, such that the models exhibit infinite towers of asymptotically conserved charges. So, further investigations are needed to uncover the bulk-boundary correspondence in the Boussinesq systems incorporating quasi-integrable deformations of the potential KdV \cite{pkdv1}. 


\noindent {\bf Acknowledgements}

We thank A. C. R. do Bomfim, L. F. dos Santos, M. Iglesias, E. Rojas and C. Murga  for useful discussions.  

\appendix 
  
\section{Potential KdV and topological solitons}
\label{app:pkdv}

The pKdV equation  (\ref{pkdv}) possesses the next kink-type solution 
\br
\label{qxt}
q(x,t) &=& q_0 + b t + a x + A_0 \tanh{[k x - \Omega t + \d_o]}.\\
b &=& - a - \frac{3}{4} \a a^2,\,\,\,
\O = \frac{1}{6} k(6  + 9 a \a  + 4 \b k^2),\,\,\,
A_0 = \frac{4 \b k}{3 \a}.
\er 
The Boussinesq $w$ and topological $Q_o$ fields can be written as functionals of the KdV $\eta$ field ($\eta = \pa_x q$)
\br
\label{wn}
w &=& - \frac{2 a k (3 a + 2 A_0 k)\b}{3 A_0} + (1+ \frac{4 a k \b}{A_0} + \frac{4 k^2 \b}{3}) \eta - \frac{A_ 0 \a + 8 k \b}{4 A_0} \eta^2.\\
\label{Qon}
Q_o &=& q_o + b t + a x \pm \frac{1}{3} \sqrt{\frac{a + A_0 k - \eta}{A_0 k}}\, (3 A_0 + 2 a k \b - 2 k \b \eta).
\er
So,
\br
\label{Qox}
\pa_x Q_o = \frac{1}{3 A_0}\Big[- 2 a k \b (3 a + 2 A_0 k) + (3 A_0 + 4 k \b (3 a + A_0 k)) \eta - 6 k \b \eta^2\Big].
\er
The above expressions satisfy the identity $\chi_{10}^2 + \chi_{20}^2 = \pa_x Q_o$ in (\ref{chi12qo}) for the bound states (\ref{chi120}).

\end{document}